\documentclass[journal]{IEEEtran}
\ifCLASSINFOpdf
\usepackage[pdftex]{graphicx}
    \setkeys{Gin}{width=.5\textwidth}
\else
\fi
\usepackage{url}

\usepackage{siunitx}

\begin{document}
\sloppy
%
\title{Measuring the Gain of a Micro-Channel Plate/Phosphor Assembly Using a Convolutional Neural Network}
%
%
%

\author{Michael~Jones,~Matthew~Harvey,~William~Bertsche,~Andrew~James~Murray,~and~Robert~B.~Appleby
\thanks{M. Jones, M. Harvey, W. Bertsche, A. J. Murray, and R. B. Appleby are with The University of Manchester, Oxford Rd, Manchester M13 9PL, UK.
M. Jones, W. Bertsche, and R. B. Appleby are with The Cockcroft Institute, UK.
Email: robert.appleby@manchester.ac.uk. }
\thanks{\textcopyright 2019 IEEE.  Personal use of this material is permitted.  Permission from IEEE must be obtained for all other uses, in any current or future media, including reprinting/republishing this material for advertising or promotional purposes, creating new collective works, for resale or redistribution to servers or lists, or reuse of any copyrighted component of this work in other works.}}

%
%

\markboth{IEEE Transactions on Nuclear Science}%
{Jones \MakeLowercase{\textit{et al.}}: Measuring the Gain of a Micro-Channel Plate/Phosphor Assembly Using a Convolutional Neural Network}
%



\maketitle

\begin{abstract}
This paper presents a technique to measure the gain of a single-plate micro-channel plate (MCP)/phosphor assembly by using a convolutional neural network to analyse images of the phosphor screen, recorded by a charge coupled device. The neural network reduces the background noise in the images sufficiently that individual electron events can be identified. From the denoised images, an algorithm determines the average intensity recorded on the phosphor associated with a single electron hitting the MCP. From this average single-particle-intensity, along with measurements of the charge of bunches after amplification by the MCP, we were able to deduce the gain curve of the MCP. 
\end{abstract}


%
\IEEEpeerreviewmaketitle

\section{Introduction}
%
%
%
%
\IEEEPARstart{M}{icro-channel} plate/phosphor assemblies comprised of a micro-channel plate and a phosphor screen are used to amplify and detect charged particle bunches in a wide variety of applications. Examples of such applications include positron detection \cite{Andresen}, X-ray spectroscopy \cite{Grantham} and ultra-fast electron diffraction experiments~\cite{VanOudheusden2010a}. In each application the total charge incident on the MCP is an important experimental parameter. In this paper we present a new technique to reliably determine the incident charge from images of the phosphor screen and measure the gain curve of an MCP.

Previous work has demonstrated how the incident charge can be determined by first measuring the gain (typically $\num{\sim1E4}$) and detection efficiency, and then recording the total charge after the bunch has been amplified. Examples of such techniques include: Oberheide \textit{et al.} \cite{Oberheide1997}, who used a photoelectron/photoion coincidence technique; Gao \textit{et al.} \cite{Gao1984} who alternatively directed ions to an MCP and a Faraday cup; and Leinard \textit{et al.} \cite{Lienard2005} who placed a transmission grid in the path of an incident ion beam to continuously measure the ion beam current.

To recover the incident charge from images of the phosphor screen, however, the user must determine the average fluorescence produced by the screen for each particle detected and amplified by the MCP. We have developed a novel technique to determine the recorded fluorescence intensity when an MCP/phosphor assembly images a single electron, which we define as the \textit{single-incident-electron-event fluorescence} (1ef). The 1ef can then be used to recover the total incident charge from images of the phosphor screen by dividing the total pixel-intensity in the images by the 1ef, after subtracting any background offset. 

Our new technique, which we present here, uses a convolutional neural network (CNN) \cite{lecun1998gradient} to remove the noise from images of the phosphor screen, thereby allowing for a reliable recovery of the 1ef for an MCP/phosphor assembly. We then demonstrate how the 1ef can be used to determine the gain of an MCP over a range of amplification bias voltages, replicating the gain curve found by Wiza \cite{Wiza1979}. We expect that these techniques will have wider application in other areas that require charged particle detection. 

\section{Experimental set-up}
To obtain images of the electron beam we use a bunched beam from an alternating current magneto-optical trap \cite{Harvey2008} cold atom electron source \cite{jones2019}. The electron bunches incident on the MCP/phosphor have a beam energy of $1.1\mbox{ keV}$. The front plate of the MCP is grounded and two bias voltages are applied to the MCP/phosphor assembly:~$5\mbox{ kV}$ with respect to ground is applied to the phosphor screen; and the bias voltage on the back plate of the MCP, which controls the MCP's gain, is varied between $0.8\mbox{ kV}$ and $1\mbox{ kV}$ with respect to ground. 


The fluorescence produced by the amplified bunch hitting the phosphor screen is recorded by a camera through a fast macro lens system. A light-tight shroud stops background light from reaching the charged coupled device in the camera. The camera has an exposure time of \SI{60}{\micro\second}, which is synchronised with the arrival of the electron bunches at the MCP.

\begin{figure}
    \centering
    \includegraphics{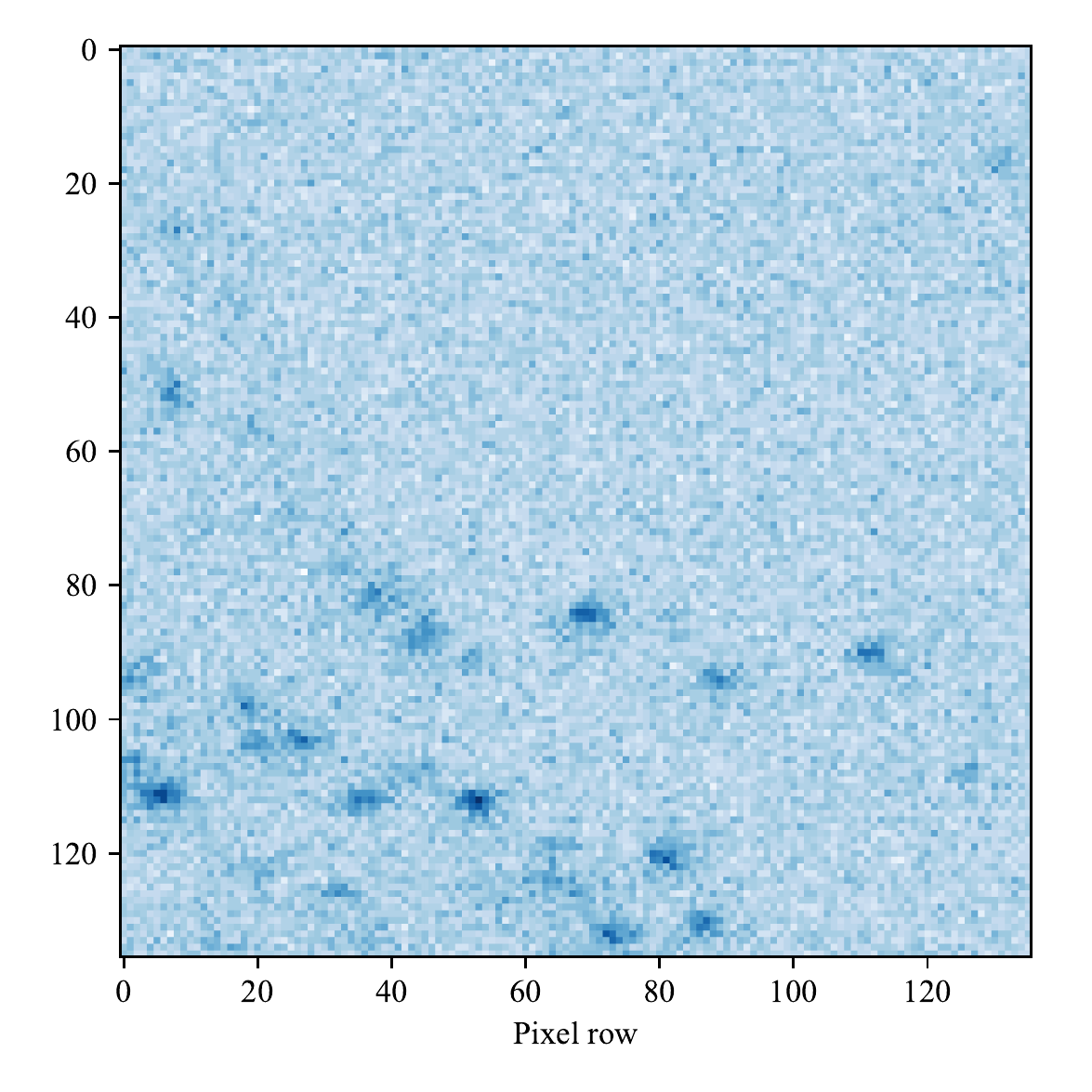}
    \caption{A typical image of a low-charge electron bunch used to estimate the 1ef. The spots in the image correspond to individual electrons imaged by the MCP/phosphor. This image corresponds to an area on the phosphor screen of \SI{2.47}{\milli\meter}~$\times$~\SI{2.47}{\milli\meter}, and was recorded for an MCP bias voltage of \SI{900}{\volt}.}
    \label{fig:electron_beam_x-sec}
\end{figure}

For each MCP back plate bias voltage, the current in the incident electron beam was reduced until single-electron events could be distinguished in the images of the phosphor. \num{250} images were recorded and \num{\sim20} clearly distinguishable spots are visible in each image. A typical image is shown in figure \ref{fig:electron_beam_x-sec}. The spots can be identified as being associated with only one electron due to both their spatial distribution in the image, and from the MCP gain curve measured using the 1ef, as discussed in section \ref{sec:det_gain}. A `background' signal was also determined by switching off the electron source and taking the average of 200 images of the phosphor screen.

\section{Determining the fluorescence when a single charged particle is imaged by the MCP/phosphor}

To deduce the 1ef a CNN first removes the noise from `denoises' the images \cite{CeLiu2008,Xie}. After denoising, each spot in the image could be reliably isolated, allowing a simple algorithm to determine the total recorded fluorescence associated with that spot.

\begin{figure}[t]
    \centering
    \includegraphics{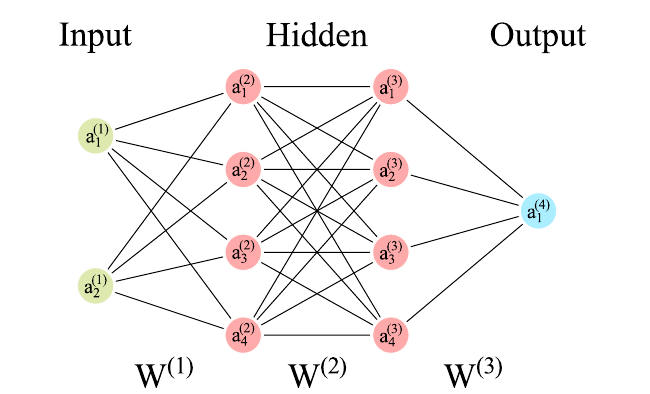}
    \caption{A typically neural network represented as a graph. The `activation', $a_i^{(n)}$ of each node is passed to all the nodes it is connected to in the next layer of the network. The connections each have an associated `weight', $w_i^{(n)}$, and the set of weights together make up the matrix $W^{(n)}$. Each node also has an associated bias, $b_i^{(n)}$. The superscript $n$ denotes which layer the node, bias, or weight in question belongs to. The input data is passed to the 'input' nodes and sets $a_i^{(1)}$, the `hidden' layers process the data, and the result is output as the activation of the 'Output' layer, $a_1^{(4)}$.}
    \label{fig:neural_network}
\end{figure}

A CNN consists of a multitude of `neural nodes', each of which takes an array of values, operates upon them, and then outputs a single value. The nodes are arranged into multiple layers, as shown in figure \ref{fig:neural_network}. The CNN which denoised the images of the phosphor screen operates on the image pixel by pixel: The CNN takes a small, 17-pixel ($17\times17$-pixel) region around a single pixel in the input image (the `receptive field') and aims to set the intensity of the corresponding pixel in the output image. The intensities of the pixels in the receptive field set the activation of the nodes in the network's first layer. The activation of the nodes in the second layer of the network is then the sum of \begin{equation}
    a_i^{(2)}=f\left(w_i^{(1)} a_i^{(1)} - b_i^{(2)}\right)
\end{equation} over all the connections to the node in question. $f(x)$ is the rectifier function, which zeroes any negative activations thereby improving the ability of the network to identify any features in the image \cite{nair2010rectified}. The nodes in each layer are connected to nodes in the next layer, and so on, until the final output layer is reached which contains a single final node. The final node's activation sets the intensity of the corresponding pixel in the output image. By optimising the weights and biases using a `training' routine, discussed below, the neural network will `learn' to reliably denoise the images~\cite{Loshchilov2017}. 

In a fully connected neural network, every node on one layer of the network is connected to every node on the next layer. The activations of the nodes on one layer are therefore the sum over all the nodes and weights from the previous layer, along with the bias associated with the node itself. For multi-layer neural networks, the number of variables which need to be optimised quickly becomes unmanageable ($\gg10,000$), particularly when dealing with large data sets, such as the 2D image data shown here. 

The increase in the number of variables when dealing with 2D image data is avoided in CNNs by exploiting the fact that the intensity of an individual pixel is likely to be dependent on the intensity of nearby pixels, and is unlikely to be related to the intensity of distant pixels. This is a reasonable assumption as nearby pixels are more likely to be part of the same feature within the image \cite{Loshchilov2017}. The 2D data can therefore be subdivided into smaller regions and each region is fed to copies of the same set of nodes, reducing the number of weights and biases which need to be managed. Such layers of the network, where sub-regions of the output from the previous layer are fed to multiple copies of the same node, are called `convolutional layers'. Another innovation in CNNs is the use of `max-pooling layers'. The exact location of an identified feature in an image is often far less critical than information about the feature's existence - for our purposes the presence of the spot corresponding to a single electron is more important than where that spot is exactly located. Therefore, taking the maximum of features over small regions in the output of a previous layer is an effective way of reducing the number of variables which must be optimised whilst retaining the ability of the network to recognise features.

The CNN used to denoise the electron images from our experiment has a structure modelled after the network designed by LeCun \textit{et al.} \cite{lecun1998gradient} and is comprised of five layers. The first layer is a convolutional layer with ten nodes. Copies of the ten nodes each take every 5-pixel sub-region (kernel) of the input 17-pixel receptive fields and processes them according to the trained weights and biases. The set of outputs is sent to a max-pooling layer which finds the maximum values of the output of the convolutional layer, pooled in a $2\times2$ window, thereby reducing the number of input nodes into the next layer by a factor of 4. The max-pooled activations are sent on to a second convolutional layer with 20 nodes and, again, a 5-pixel kernel. The output from the second convolutional layer is fed into a fully connected layer with ten nodes which, in turn, feeds into the final single node. The final node stores the intensity of the pixel in the resulting denoised image.

\begin{figure}
    \centering
    \includegraphics{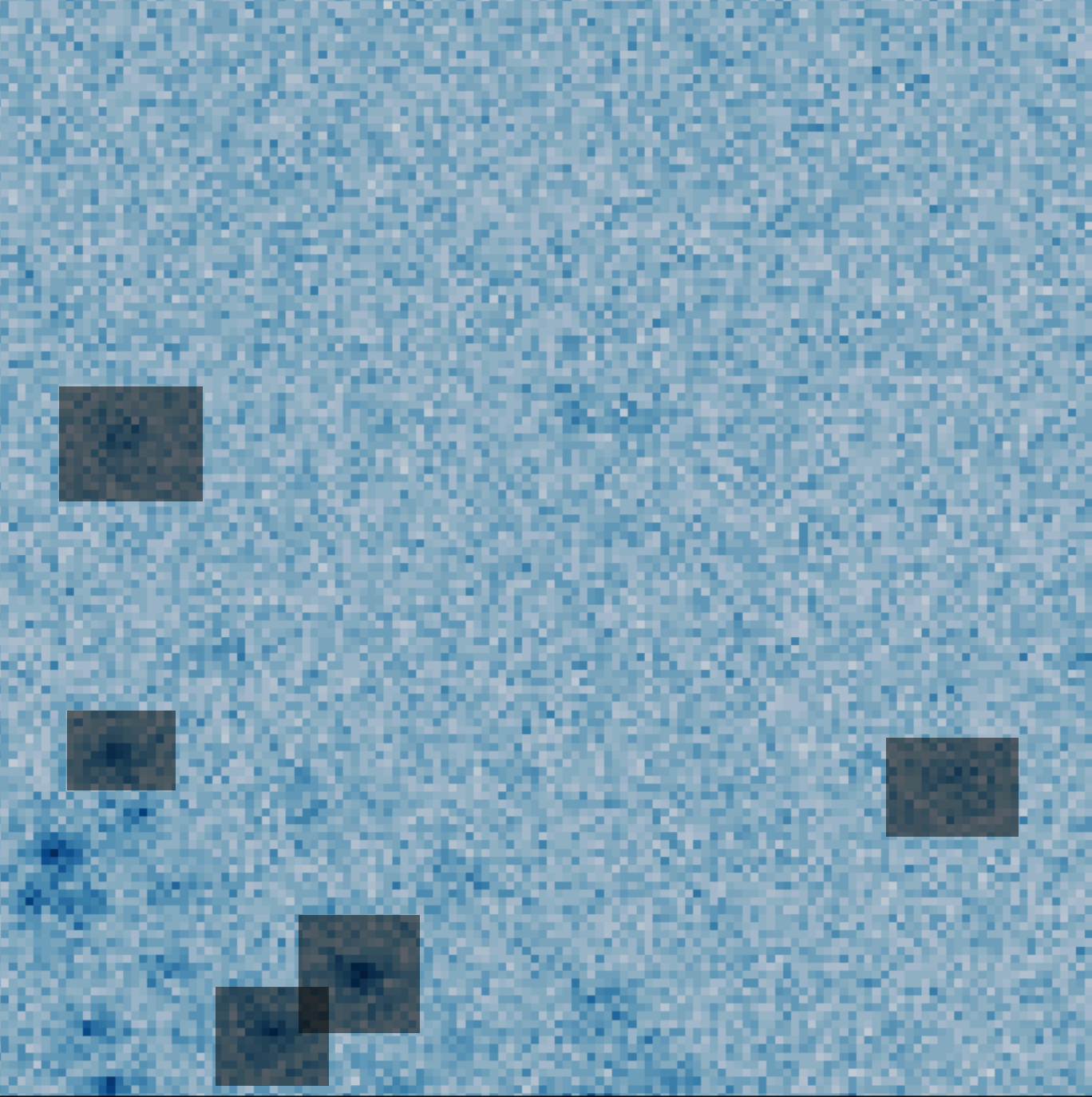}
    \caption[Example of a single image in the process of being tagged]{Example of an image in the process of being tagged using the JavaScript application. The dark regions are electrons that have already been tagged. }
    \label{fig:tagged_image}
\end{figure}
\begin{figure}
    \centering
    \includegraphics{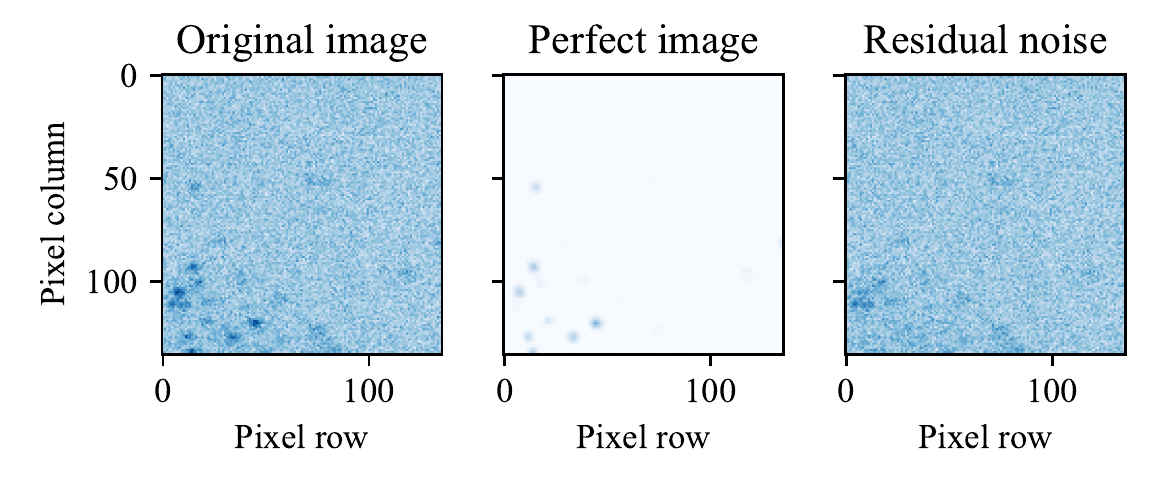}
    \caption[An original image, a perfect image, and the residual noise]{The original image and the image created by fitting a 2D Gaussian to each of the tagged regions, with the residual noise.}
    \label{fig:perfect_image}
\end{figure}

The CNN is trained to denoise electron images using a set of training data consisting of noisy images and their 'perfect', noiseless, counterparts. To produce the training data approximately 1000 single electron spots in the first 50 images in the series were `tagged' using a browser-based application. 
An example of an image in the process of being `tagged' in the application is shown in figure \ref{fig:tagged_image}. Once the 1000 electrons are tagged, a Python script fits a 2D Gaussian to each of the tagged regions to approximate the image of the underlying spot without any noise. The set of noiseless spots from each image are then recombined into a series of `perfect' images, an example of which is shown in figure \ref{fig:perfect_image}. By allowing the user to discriminate against clustered events and overlapping electron spots in the tagging process, our procedure reduces the skew in the 1ef distribution caused by these processes.

Every pixel in the noisy images, along with the 17-pixel receptive field which surrounds it, is paired with a corresponding pixel in the `perfect' images to create the set of training data. The CNN is trained using a quasi-Newton stochastic gradient descent (SGD) algorithm called Adam~\cite{Loshchilov2017}. First Adam selects an initial set of weights and biases for the network using a Xavier initialisation routine \cite{DelBue2010} which assigns initial weights and biases according to a Gaussian distribution with a mean of zero and a finite variance. A randomly selected subset of the training data is then passed through the network. Adam then defines a loss function, $C(w_i,b_i)$ as the sum of squared differences between the `perfect' pixel values and the output of the network, over the selected subset of the training data. This loss function is used to judge the performance of that set of weights and biases. Adam calculates the gradient of the loss function, $\nabla C$ for the initial set of weights and biases, and takes a step in the direction of $-\nabla C$. After many iterations, Adam will reliably reach a local minimum of the loss function \cite{Loshchilov2017}. That set of weights and biases produces a neural network which can predict the pixel values in the perfect output images from the pixel values in the input images.

It is important to note that since only a randomly selected subset of the training data is used to approximate the gradient of the loss function in each step of the SGD algorithm, the approach to a local minimum of the loss function is stochastic. We use the SGD algorithm instead of a direct gradient descent algorithm using the entire set of training data since the SGD algorithm is computationally faster. The stochastic nature of the SDG algorithm also reduces the probability of the descent being trapped in a local minimum farabove the global minimum of the loss function. The neural network and the training algorithm were implemented using the Python library PyTorch \cite{paszke2017automatic}.

\begin{figure}
    \centering
    \includegraphics{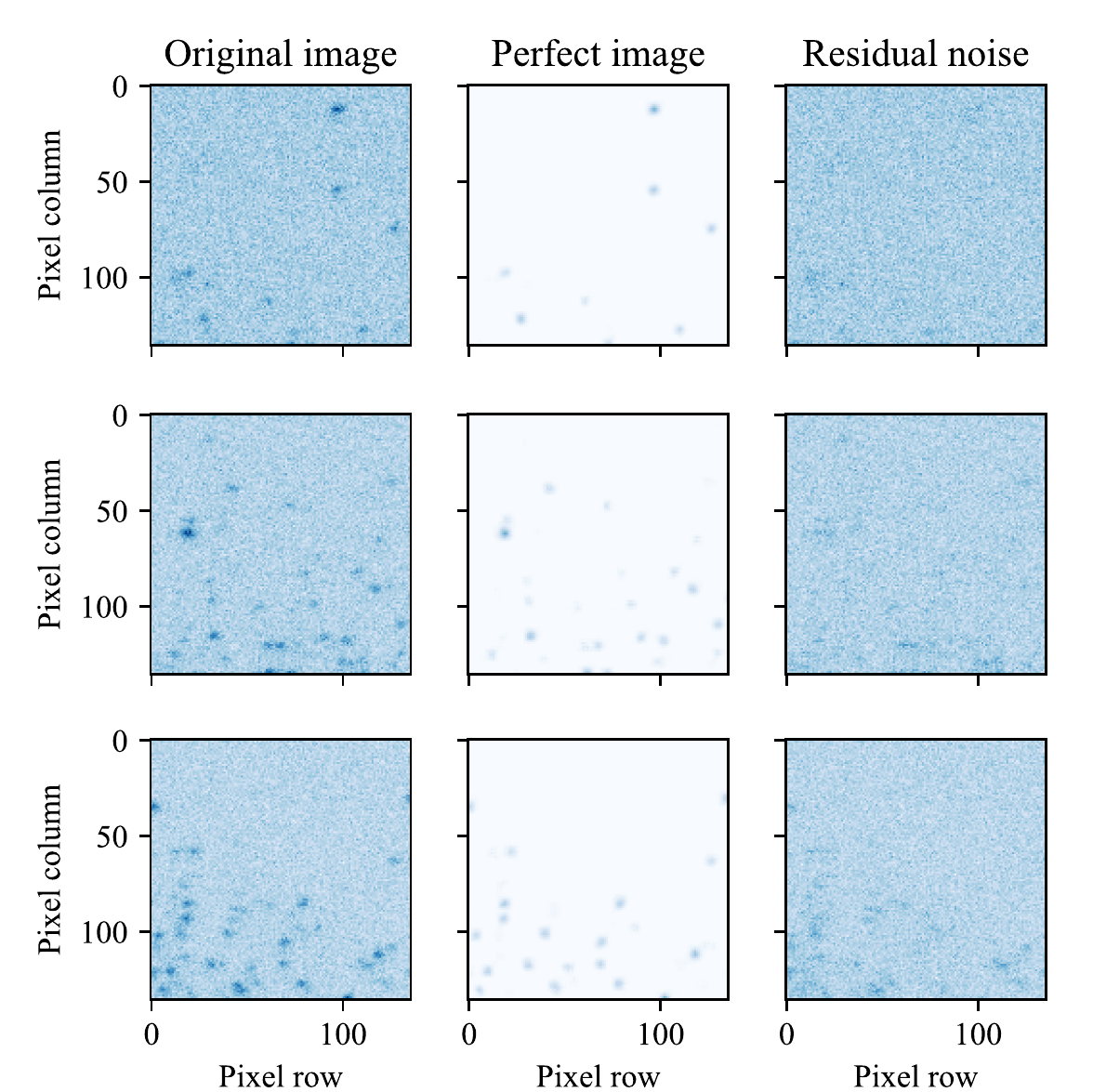}
    \caption[An original image, a `perfect' image calculated by the neural network, and the residual noise]{A series of raw images and the corresponding `perfect' images as calculated by the neural network. The residual noise after subtracting the perfect image from the original image is also shown. Electron signals with significant overlap with one another are ignored by the neural network as they are ignored during the tagging process, and are therefore not present in the `perfect' set of images used to train the network. Hence, the indistinct electron spots remain in the `residual noise' images.}
    \label{fig:nn_output}
\end{figure}

\begin{figure}
    \centering
    \includegraphics{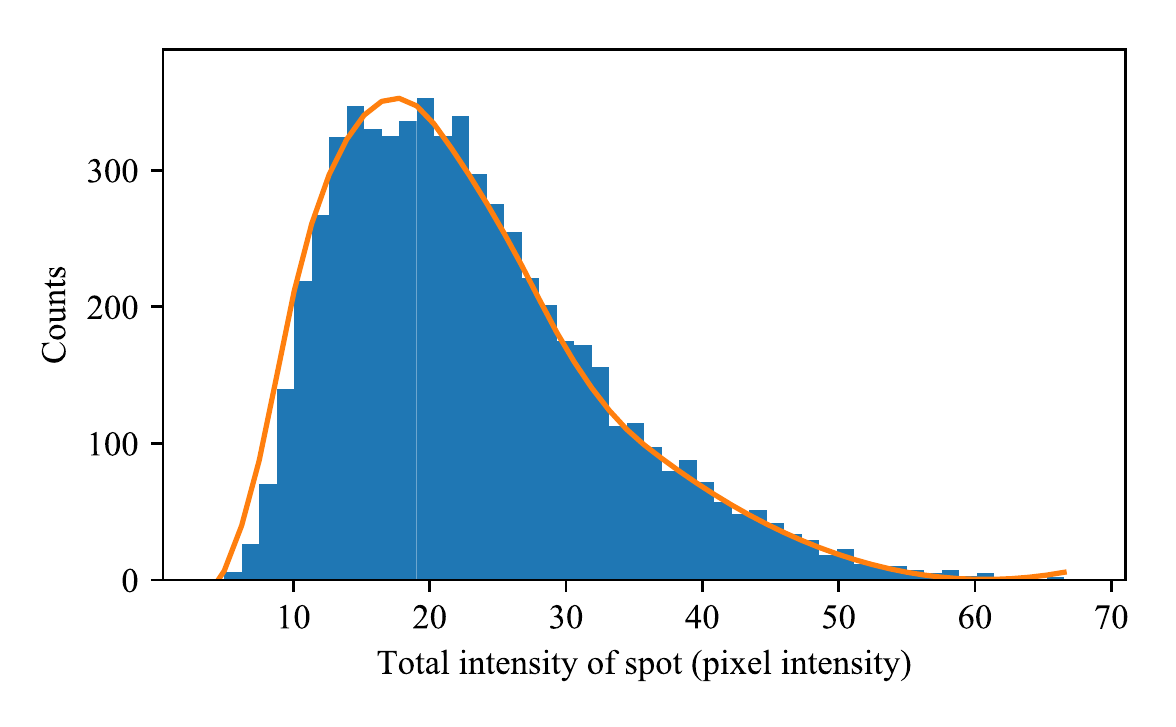}
    \caption[Single electron intensity distribution]{The single electron intensity distribution calculated by summing the intensities of the pixels which make up each spot in the noise-free, overlap-free, images. The bias of the distribution originates from the few overlapped electrons which escape the denoising process and are present in the final `perfect' images, hence we use the peak of a spline fitted to the distribution (orange) as our measure of the average 1ef. }
    \label{fig:1e_intensity}
\end{figure}

The trained neural network generates a set of `perfect' images based on the full set of noisy input images. Examples of the input and corresponding output images along with the residual noise are shown in figure \ref{fig:nn_output}. A threshold function can then separate each single electron event in the output images from the others as the stochastic noise in the images is so low, allowing us to deduce the total pixel-intensity attributable to each event. From the distribution of those $\sim3200$ measurements, shown in figure~\ref{fig:1e_intensity}, we could establish the modal intensity of a single electron event as \num{19.8(10)} units of pixel intensity. We use this modal value to derive the bunch charge of the electron beam from the associated cross-section images. 

\section{Determining the MCP gain from the recorded fluorescence distribution}\label{sec:det_gain}

\begin{figure}
\centering
    \includegraphics{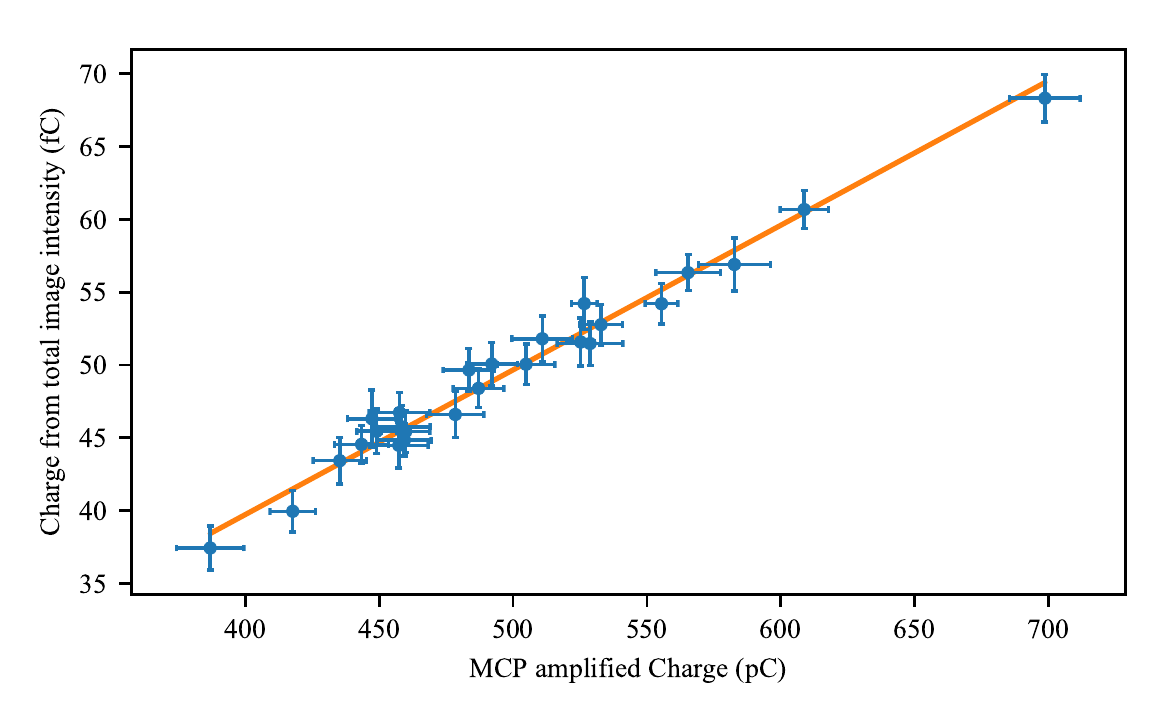}
    \caption{Comparing charge from the MCP to the sum of the pixel values recorded by the camera for 100 electron bunches at varying bunch charges shows the linear relationship between the two values. The gain of the MCP can be determined from the relationship between the single electron intensity distribution, the total image intensity, and the bunch charge amplified by the MCP.}
    \label{fig:MCP_gain_int}
\end{figure}

Finding the single electron intensity allows us to calculate the gain of the MCP as the ratio of charge in the incident electron bunch, determined from the images of the phosphor, to the total charge produced by the MCP. The gain of the MCP was determined in two stages. First, the MCP was used to amplify a series of `high charge' electron bunches which were then imaged by the phosphor screen. We plotted the total image intensity (after background removal) against the integral of the current from the front plate of the MCP for a range of bunch charges. The plot reveals a linear relationship between image intensity and bunch charge which is shown in figure~\ref{fig:MCP_gain_int}, establishing the linearity of the MCP over the range of bunch charges.

Dividing the total intensity of an image of an electron bunch by the mean single electron intensity gives a value for the charge of the bunch that produced the image, without requiring that the electron events in the image are individually distinguishable. Plotting the intensity-derived figure for the bunch charge against the total amplified charge from the MCP then gives a value for the gain of the MCP of \SI{9.15(11)E3}{} for an MCP bias voltage of \SI{900}{\volt}. This figure is consistent with the figure from the literature of \num{\sim9.2E3} \cite{Wiza1979}. The agreement of our gain calculation with the literature figure shows the efficacy of using the CNN technique to determine the single electron intensity distribution. 

\begin{figure}
\centering
    \includegraphics{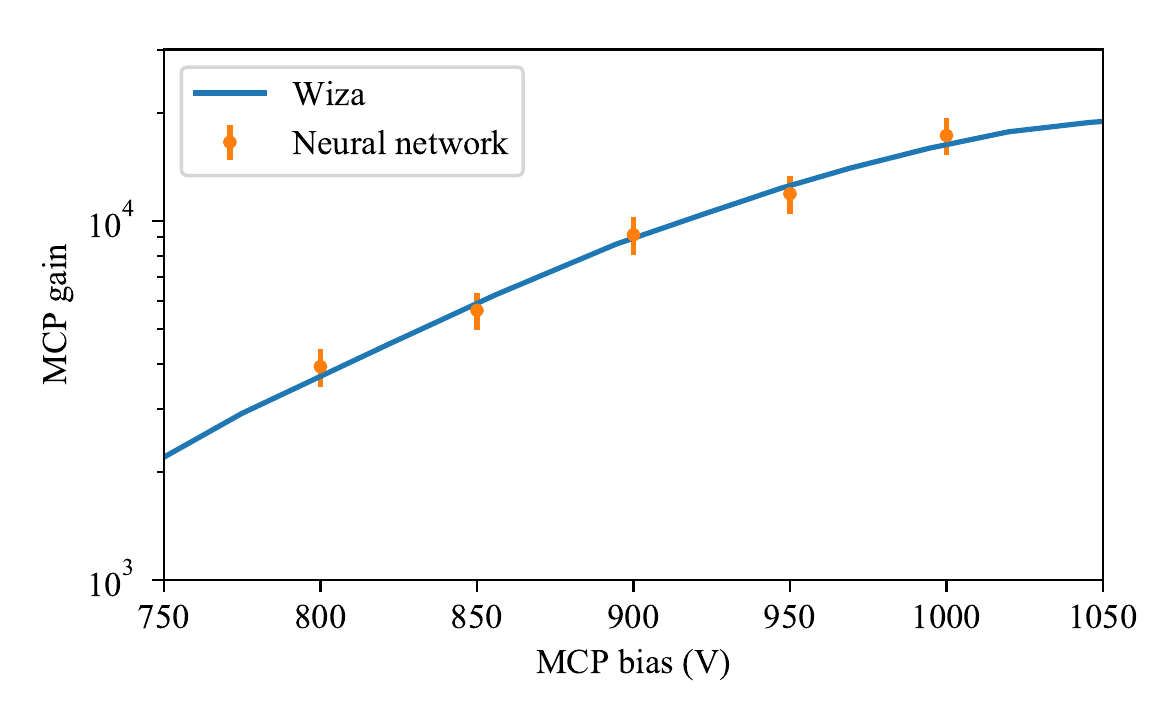}
    \caption{The gain curve for the MCP over a range of bias voltages. The gain curve calculated using the neural network is consistent with the gain curve from Wiza \cite{Wiza1979}, shown as the blue line.}
    \label{fig:gain_curve_MCP}
\end{figure}

We repeated the above analysis for the images taken over a range of MCP bias voltages. The resulting gain curve, for electrons at \SI{1.1}{\kilo\electronvolt}, as a function of the MCP bias voltage is shown in figure~\ref{fig:gain_curve_MCP}. The gain curve calculated using the neural networks shows excellent agreement with that measured by Wiza~\cite{Wiza1979}.

\section{Conclusion}
We have demonstrated that the gain curve of an MCP/phosphor assembly can be determined using images of low density electron bunches that are amplified and imaged by the assembly, and a convolutional neural network. This technique has applications where electronic noise makes other methods of MCP gain calibration, such as using a Faraday cup or a transmission grid, very difficult. By measuring the charge using the images of the phosphor we were able to reduce the complexity of our experiment, as we no longer needed to record the charge data separately from the image data.

Finally, characterising the fluorescence associated with single electron events allowed us to produce well parameterized `pseudo-data' with which we could test other image analysis techniques. More details of this process are given in \cite{jones2019}. Our code, created to tag and analyse the images and extract the 1ef, is hosted at \url{https://github.com/mikeedjones/denoising}, along with further documentation detailing its use.


%



\section*{Acknowledgment}

The authors would like to thank Mark Surman of the Cockcroft Institute for the generous lending of equipment, without which this work would not be possible. A. J. Murray and M. Harvey would also like to acknowledge the EPSRC for grant R120272 and all authors
acknowledge the Cockcroft Institute and STFC through grant number ST/P002056/1.

\ifCLASSOPTIONcaptionsoff
  \newpage
\fi



%

\bibliographystyle{IEEEtran}
\bibliography{bib/references}

\vfill


\end{document}